\documentclass[twocolumn,showpacs,preprintnumbers,amsmath,amssymb]{revtex4}

\usepackage{graphicx}
\usepackage{dcolumn}
\usepackage{bm}

\begin{document}


\title{Electron-beam propagation in a two-dimensional electron gas}

\author{E.~G.~Novik, H.~Buhmann, and L.~W.~Molenkamp}
\affiliation{ Physikalisches Institut der Universit\"{a}t
W\"{u}rzburg, Am Hubland, 97074 W\"{u}rzburg, Germany }

\date{\today}

\begin{abstract}
A quantum mechanical model based on a Green's function approach
has been used to calculate the transmission probability of
electrons traversing a two-dimensional electron gas injected and
detected via mode-selective quantum point contacts.
Two-dimensional scattering potentials, back-scattering, and
temperature effects were included in order to compare the
calculated results with experimentally observed interference
patterns. The results yield detailed information about the
distribution, size, and the energetic height of the scattering
potentials.
\end{abstract}


\maketitle

\section{\label{sec:level1}Introduction}

The electron mobility in a two-dimensional electron gas (2DEG) is
limited by electron-impurity scattering \cite{ref1,ref2,ref3}. The
impurity potential originates mainly from ionized donors located
in the remote doping layer. However, the observed high mobilities
can only be explained when correlations of different donor states
are taken into account \cite{ref4,ref5}. Up to now experiments are
rare which probe the actual impurity potential microscopically.
Standard transport experiments yield only information about the
averaged potential fluctuations by defining the average electron
mean free path. Recently, it has been shown that the electrons
injected via quantum point contacts (QPC) into a 2DEG can be used
as a local probe for these scattering potentials either through
the interference patterns observed in the transmission probability
\cite{ref6} or through structures in the position dependent
electron reflection \cite{ref7,ref16}.

In this paper we use a Green's function approach to calculate the
transmission probability of electrons injected via QPC into a 2DEG
exposed to a small magnetic field. The results are compared with
experimentally observed transmission probabilities where the
electrons were detected with a second QPC in a distance of
4~$\mu$m opposite to the injector QPC. In order to reproduce the
experimental results the shape, size, and height of scattering
potentials has been modeled and back-scattering as well as
temperature effects have been included into these calculations.

\section{\label{sec:level1}Model}

Our model is adopted to the experimental situation presented in
Fig.~1~a) (c.f. Ref.~\cite{ref6}). This figure shows two opposite
QPCs in a distance $L$ from each other. In the experiment these
QPCs are formed electrostatically in the 2DEG of
GaAs/Al$_{0.33}$Ga$_{0.67}$As-heterostructure by externally
controlled Schottky gates (gray areas). Due to the
saddle-point-like electrostatic potential of the QPCs in the plane
of the 2DEG the electrons are injected into the region between the
two QPCs in form of a collimated beam \cite{ref8}. A weak magnetic
field, applied perpendicular to the plane of the 2DEG, deflects
the electrons and the resulting signal of the electrons which
reach the detector QPC is proportional to the profile of the
propagated electron beam. It has been shown in Ref.~\cite{ref6}
that the observed structure in the electron beam signal is due to
electron interference effects, which originate from scattering at
potential fluctuations imposed into the 2DEG from donors in
different charged states. A typical experimental result is shown
in Fig.~\ref{fig2} (curves E).

The first attempt to calculate the transmission probability of an
electron beam injected by a QPC into a 2DEG and detected by a
second opposite QPC was made by M.~Saito \textit{et al.}
\cite{ref9} using a Green's function approach. This approach was
previously extended by us in order to include interference effects
due to electron impurity scattering \cite{ref6}. The scattering
potential has been treated in a very simplified way by setting the
part of a wave function at the impurity position to zero (infinite
impurity potential), which explained the experimental data
qualitatively. Here, we present quantum mechanical calculations
which include a realistic two-dimensional impurity potential model
as well as back-scattering and temperature effects. A fitting to
the experimental data yields information about the distribution,
size and strength of scattering potentials in a 2DEG. In contrast
to our previous work \cite{ref6}, finite potential fluctuations
with positive and negative heights (counted from the bottom of the
conduction band) are considered, corresponding to the regions of
reduced and increased electron density, respectively. Moreover,
the back-scattering effects from the impurity potentials and
sample boundaries were taken into account. It turns out that the
back-scattering affects the interference patterns considerably in
two cases: first, if the impurity is located close to the injector
or the detector QPC (within the phase coherence length), or
second, if two impurities are located at a short distance from
each other. The effect is stronger when the scattering potential
is significantly higher than the Fermi energy. Finally,
temperature effects are considered, i.e., a thermal averaging of
the propagated beam is included \cite{ref10}, as well as the
contribution from electron-electron scattering \cite{ref11}.

\begin{figure}
\includegraphics[width=8cm]{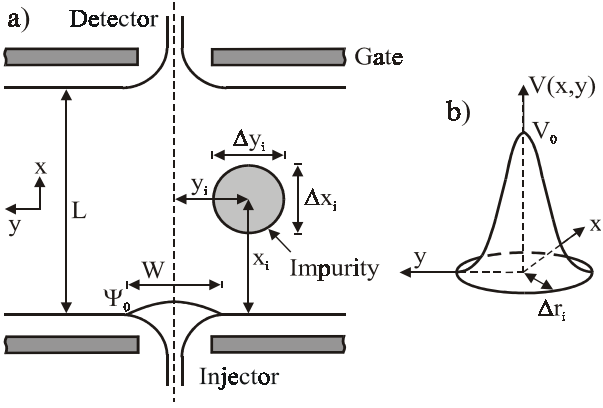}
\caption{\label{fig1} a)~Scheme of the sample structure used for
the calculations ($\Psi_{0}$~- the wave function at the exit of
injector quantum point contact); b)~the shape of the scattering
potential.}
\end{figure}

\begin{figure}
\includegraphics[width=8cm]{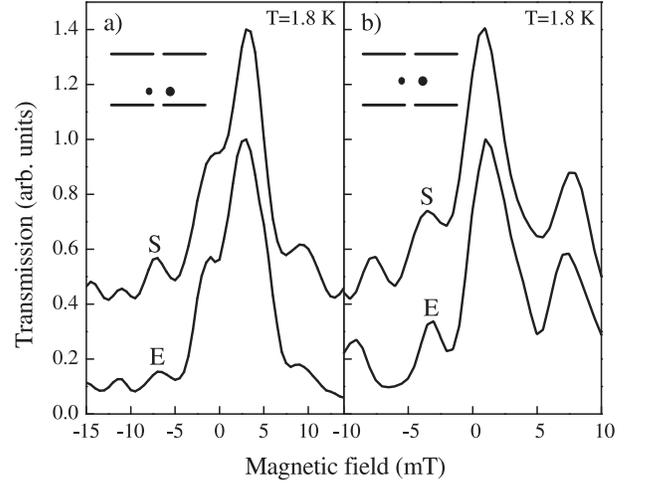}
\caption{\label{fig2} Measured electron beam profiles (curves E)
and calculated interference patterns (curves S) for two scattering
potentials versus magnetic field (simulation results are presented
with a shift). Parameters for the scattering potentials are:
a)~$x_i=1.3~\mu$m, $y_i=-0.34~\mu$m, $\Delta r_i=0.045~\mu$m,
$V_{0}\approx33$~meV for first impurity and $x_{i}=1.3~\mu$m,
$y_i=0.18~\mu$m, $\Delta r_i=0.075~\mu$m, $V_{0}\approx16$~meV for
second; b)~$x_i=2~\mu$m, $y_i=-0.42~\mu$m, $\Delta r_i=0.05~\mu$m,
$V_{0}\approx8$~meV for first impurity and $x_i=2~\mu$m,
$y_i=0.215~\mu$m, $\Delta r_i=0.075~\mu$m, $V_{0}\approx10$~meV
for second.}
\end{figure}

The following initial parameters are set according to the
experimental situation for the calculations: the distance ($L$)
between the two QPCs is $4~\mu$m, which is smaller than the
elastic mean free path, the electron mobility and carrier density
are $\mu\approx100$~m$^{2}/$Vs and
$n_s\approx2\times10^{15}$~m$^{-3}$, respectively, and the width
($W$) of the QPC's exit is 100~nm accounting for a single mode
electron injection in case of an adiabatic expansion
\cite{ref6,ref8}. The wave function propagation is perturbated by
potential fluctuations, which are described in the following way:
$x_{i}$, $y_{i}$, and $\Delta x_{i}$, $\Delta y_{i}$ define the
position and extension in the $x$-$y$ plane and $V(x,y)$ the
potential height (Fig.~\ref{fig1}). It was found that in most
cases circular impurities ($\Delta x_{i}=\Delta y_{i}$) are
sufficient to produce a good agreement between the calculations
and the experiments. Hence, we restrict ourself in this paper to
the presentation of results considering only circular impurity
potentials with a radius which is given by $\Delta
r_{i}=\sqrt{\Delta x_{i}^{2}/4+\Delta y_{i}^{2}/4}$.

Our model numerically solves the time-independent Schr\"{o}dinger
equation for a static magnetic field \cite{ref9}. Using the
Green's function method with Dirichlet's boundary conditions it is
possible to evaluate the wave function for a line at any distance
$x'$ from the injector, according to \cite{ref9}:

\begin{equation}
\Psi(x',y')=\frac{\hbar}{2m^{*}}\int\limits_{-W/2}^{W/2}\Psi_{B}(0,y)\left.\,\frac{\partial\,
G^{+}}{\partial\,x}~\right|_{x=0}dy . \label{eq-1}
\end{equation}
Here, $\Psi_{B}$ is the wave function at the injector QPC exit
which includes the effect of the magnetic field in the form of the
phase shift $\Theta_{0}$ between the center and any point on the
QPC exit:

\begin{equation}
\Psi_{B}(0,y)=\Psi_{0}(0,y)\exp[i\Theta_{0}(y)] , \label{eq-2}
\end{equation}
$\Psi_{0}$ is the wave function at the injector QPC exit without
magnetic field and $G^{+}$ is the Green's function calculated in a
weak-magnetic-field approximation using the mirror-image method.
For a detailed description of this method we refer to
Ref.~\cite{ref9}. The unperturbated wave function can be
calculated for the whole 2DEG area between injector and detector.
However, in the region $x_i-\Delta r_i<x<x_i+\Delta r_i$,
$y_i-\Delta r_i<y<y_i+\Delta r_i$ the wave function propagation is
perturbated by a scattering potential. One suitable approximation
for the shape of this potential is a hyperbolic function
\cite{ref12}, which is very similar to a Gaussian potential
commonly used in literature for the description of the effective
potential due to remote charged donors in the 2DEG region
\cite{ref13}. A hyperbolic potential has the advantage that the
solution for the transmission and reflection coefficients can be
obtained analytically \cite{ref12}. The scattering potential is
given by [Fig.~\ref{fig1}~b)]:

\begin{equation}
V(x,y)=V_{0}\cosh^{-2}\left[\frac{\sqrt{(y-y_{i})^{2}+(x-x_{i})^{2}}}{a\Delta
r_{i}}\right] , \label{eq-3}
\end{equation}
$V_{0}$  defines the height of the potential which can be positive
and negative depending on its relative height compared to the
bottom of the conduction band for the average carrier density
\cite{ref14}. The constant $a$ has been chosen $a\approx0.4$ in
order to allow for the following approximation: $V(x,y)\approx0$
for $\sqrt{(y-y_{i})^{2}+(x-x_{i})^{2}}>\Delta r_i$, i.e. the
impurity potential acts as a local perturbation.

The effect of the scattering potential is calculated in the
following way: first, the unperturbated wave function is
calculated for a line $x=x_i+\Delta r_i$ behind the impurity
potential. Then the transmission probability of this scatterer is
evaluated and the resulting wave function at $x=x_i+\Delta r_i$ is
calculated from:

\begin{equation}
\Psi_{T}(x_{i}+\Delta r_i,y)=\Psi(x_{i}+\Delta r_i,y)~t(y) ,
\label{eq-4}
\end{equation}
where $\Psi(x_{i}+\Delta r_i,y)$ is the wave function, calculated
omitting the impurity. As long as the size of the scattering
potential is much smaller than the distance between injector and
detector QPC, the transmission amplitude can be calculated from:

\begin{equation}
t(y)=\frac{\Gamma(-i\bar{k}-s(y))~\Gamma(-i\bar{k}+s(y)+1)}{\Gamma(1-i\bar{k})~\Gamma(-i\bar{k})}
, \label{eq-5}
\end{equation}
where
\begin{equation}
s(y)=\frac{1}{2}\left(-1+\sqrt{1-\frac{8m^{*}V(x_{i},y)}{(\hbar/a\Delta)^{2}}}~\right)
, \label{eq-6}
\end{equation}
and
\begin{equation}
\bar{k}=ka\Delta , \label{eq-7}
\end{equation}
where $\Gamma$ is the gamma function,
$k=\frac{\sqrt{2m^{*}E}}{\hbar}$; $V(x_{i},y)$ is the height of
the impurity at the position $y$ [Eq.~(\ref{eq-3})];
$\Delta=\sqrt{\Delta r_i^{2}-(y_i-y)^{2}}$ (where $|y-y_i|<\Delta
r_i$) is the half of the extension of the impurity in
$x$-direction for a given $y$-coordinate. Multiple impurity
potentials are considered in the same successive way. The modified
wave function is then used to calculate  the wave function at the
detector position ($x=L$) using Eq.~(\ref{eq-1}). The transmission
coefficient for the detector QPC is calculated from

\begin{equation}
T=\left|\int\limits_{-W/2}^{W/2}\Psi_{D}^{*}(L,y')\Psi(L,y')\,dy'\right|^{2}
, \label{eq-8}
\end{equation}
where $\Psi_{D}(L,y')$ is the wave function in the detector QPC,
which is taken in analogy to the wave function $\Psi_{B}$ at the
exit of the injector QPC \cite{ref6}.

From comparing the calculated transmission probability with the
measured beam profile it is possible to determine the parameters
of the scattering potential. As an example two experimental traces
and matching theoretical results are presented in Fig.~\ref{fig2}.
Good agreement is obtained for both cases when two scattering
potentials are introduced into the region between injector and
detector. The sizes of the impurities are comparable with Fermi
wavelength $\lambda_{F}\approx0.05~\mu$m, and their heights are
equal or exceed Fermi energy $E_{F}\approx8$~meV.

\begin{figure}
\includegraphics[width=8cm]{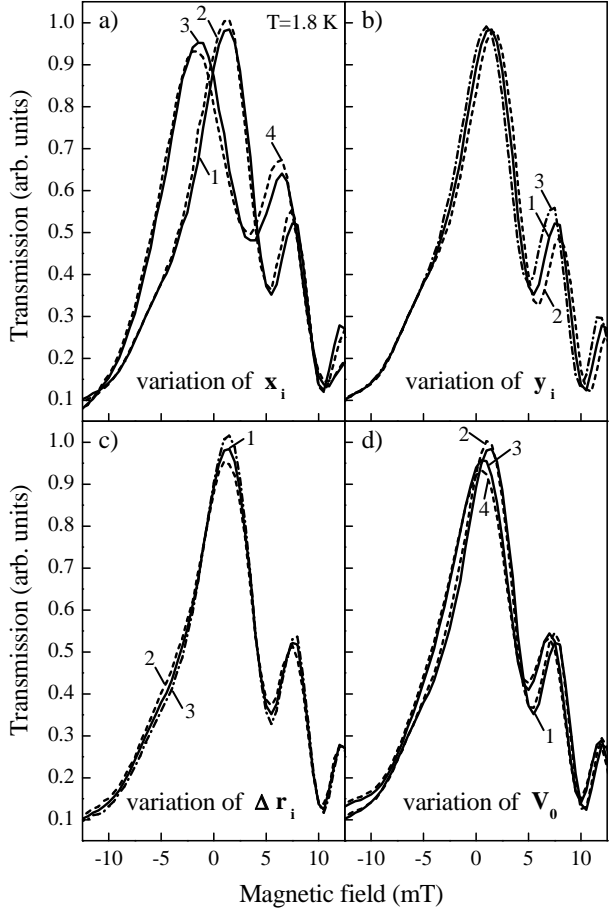}
\caption{\label{fig3} Calculated interference patterns for
different values of the impurity potential parameters. Curve~1 in
all for figures: $x_i=2~\mu$m, $y_i=0.205~\mu$m, $\Delta
r_i=0.035~\mu$m, $V_{0}\approx16$~meV; a)~$x_i=2.4~\mu$m
(curve~2), $x_i=0.6~\mu$m (curve~3) and $x_i=0.57~\mu$m (curve~4);
b)~$y_i=0.195~\mu$m (curve~2) and $y_i=0.215~\mu$m (curve~3);
c)~$\Delta r_i=0.031~\mu$m (curve~2) and $\Delta r_i=0.039~\mu$m
(curve~3); d)~$V_{0}\approx12$~meV (curve~2),
$V_{0}\approx7.3$~meV (curve~3) and $V_{0}\approx6.5$~meV
(curve~4).}
\end{figure}

\begin{figure}
\includegraphics[width=6.6cm]{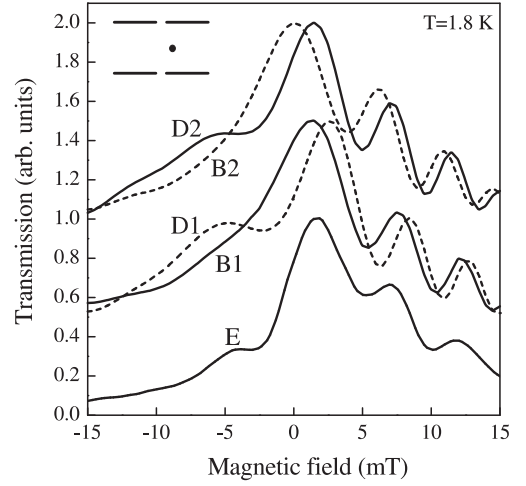}
\caption{\label{fig4} Calculated interference patterns for a
barrier (curves B1 and B2) and for a dip (curves D1 and D2) in
comparison with experimental data (curve~E). All characteristics
are normalized with respect to the maximum transmission value.
Results are displayed with a constant shift for clarity.
Parameters for the impurities are: $x_i=2~\mu$m, $y_i=0.205~\mu$m,
$\Delta r_i=0.035~\mu$m, $V_{0}\approx16$~meV (curve B1);
~$x_i=2~\mu$m, $y_i=0.205~\mu$m, $\Delta r_i=0.035~\mu$m,
$V_{0}\approx-16$~meV (curve D1); ~$x_i=2~\mu$m, $y_i=0.24~\mu$m,
$\Delta r_i=0.035~\mu$m, $V_{0}\approx14$~meV (curve B2);
~$x_i=2~\mu$m, $y_i=0.24~\mu$m, $\Delta r_{i}=0.035~\mu$m,
$V_{0}\approx-14$~meV (curve D2).}
\end{figure}

The sensitivity of the interference pattern to the parameters of
the scattering potential is demonstrated in Fig.~\ref{fig3} for
the case of a single impurity. The initial parameters (location
$x_i=2~\mu$m, $y_i=0.205~\mu$m, size $\Delta r_i=0.035~\mu$m and
height $V_{0}\approx16$~meV) were those which give the best fit of
the experimental data presented in Fig.~\ref{fig4} (curve E). In
Fig.~\ref{fig3}~a) the influence of the parameter $x_i$ on the
interference pattern is demonstrated: if the impurity is located
far away from the injector QPC (and detector), e.g. $x_i=L/2$, the
observed interference patterns are relative insensitive to small
changes in $x_i$. A change of about $20\%$ is needed to give the
result shown in Fig.~\ref{fig3}~a) trace 2. However, if the
impurity is located much closer either to injector or detector
[Fig.~\ref{fig3}~a) trace 3], changes in $x_i$ of $5\%$ result in
a similar strong effect (trace 4) as observed for the center
position. The influence of the parameter $y_i$ is strong but only
weakly dependent on the $x$-position: changes of $y_i$ by $5\%$
lead to shifts of the interference maxima by $1-6\%$ [traces 2 and
3 in Fig.~\ref{fig3}~b)]. Fig.~\ref{fig3}~c) shows the sensitivity
of the interference patterns to the variation of the parameter
$\Delta r_i$ by $10\%$. The interference maxima for curves 2 and 3
are shifted relative to the initial curve by $1-4\%$, which
demonstrates also here the relatively high sensitivity of the
fitting procedure. The influence of the parameter $V_{0}$ is quite
different depending on the absolute height compared with the Fermi
energy. Two examples are given in Fig.~\ref{fig3}~d) where the
interference patterns for $V_{0}\approx16$~meV (trace~1) and
$V_{0}\approx7.3$~meV (trace~3) are shown. The lower potential
height corresponds closely to the Fermi energy. It can be seen
that a change of $V_{0}$ of $25\%$ (trace~2) in the first case
($V_{0}>E_{F}$) is need to achieve a similar deviation from the
initial curve as it has been obtained for a change of only $10\%$
(trace~4) in the case when $V_{0}$ is comparable to $E_{F}$.

It should be noted that the simulations do exhibit an ambiguity in
the determination of the impurity position in the direction of
electron beam propagation (i.e. in the determination of the
position $x_{i}$). Because of the symmetry of the device, a
similar transmission probability results for an impurity located
at a distance $x_{i}$ or $x'_{i}\approx L-x_{i}$ from the injector
QPC boundaries, when the other impurity parameters are kept
constant. A further ambiguity is in the determination of the
character of the scattering potential, i.e., it is possible to
obtain similar interference patterns for a barrier ($V_{0}>0$) and
for a dip ($V_{0}<0$). In Fig.~\ref{fig4} the calculated
transmission probability for a potential barrier (curve~B1) and a
dip (curve~D2) are shown. The results reproduce the characteristic
features of the experimental data quite well. The scatterers have
in both cases a radius $\Delta r_i=0.035~\mu$m but they differ in
position and height. Optimal locations found in the simulation
were $x_i=2~\mu$m, $y_i=0.205~\mu$m for the potential barrier and
$x_i=2~\mu$m, $y_i=0.24~\mu$m for the dip, for potential heights
are $V_{0}\approx16$~meV and $V_{0}\approx-14$~meV, respectively.
It can be seen that a simple change of the sign of the potential
height does not result in satisfying fitting of the experimental
data. Fig.~\ref{fig4} shows the inversion from a barrier
(curve~B1) into a dip (curve~D1) and from a dip (curve~D2) into a
barrier (curve~B2). From these results it can be seen, that the
absolute value of this fitting parameter depends mainly on the
choice of the character of the scattering potential (dip or
barrier). However, for most of our calculations we decided on
barrier type potentials which correspond to 2DEG areas of reduced
carrier density. In real GaAs 2DEG structures this type of
scattering potentials is expected due to the formation of
negatively charged DX centers in the remote donor layer
\cite{ref5}.

So far we have not taken into account any temperature effects.
Experimentally the interference patterns have been observed to
smear out with increasing temperature, practically disappearing at
$T\approx4$~K [Fig.~\ref{fig5}~a)]. We first consider thermal
broadening as the origin of this behavior. The detector signal
results from electrons with different energies $E_{F}\pm k_{B}T$
propagating from injector to detector. In order to include this
effect the calculations of the wave function [Eqs.~(\ref{eq-1},
\ref{eq-4}-\ref{eq-7})] were carried out for the corresponding
values of wave vector $k=\frac{\sqrt{2m^{*}(E_{F}\pm
k_{B}T)}}{\hbar}$. The thermally averaged transmission probability
is then calculated as:

\begin{equation}
T=\left|\int\limits_{}^{}-\frac{\partial\,f(E)}{\partial\,E}\left[\int\limits_{-W/2}^{W/2}\Psi_{D}^{*}(L,y')\Psi(L,y',E)\,dy'\right]dE\right|^{2}
, \label{eq-9}
\end{equation}
where $-\frac{\partial\,f(E)}{\partial\,E}$ is the derivative of
the Fermi function \cite{ref10}. From Fig.~\ref{fig5}~b) it can be
seen that the reduction of the amplitude of the detector signal
with increasing temperature is reasonably reproduced by the
thermal broadening. However, the signal still shows all features
corresponding to the low temperature interference patterns
[Fig.~\ref{fig5}~b) solid lines], while experimentally these
features are all smeared out. This result implies that further
temperature effects are responsible for the experimental
observations. Another phenomenon which strongly influences the
appearance of interference effects in an electron system is
electron-electron scattering. It leads to a dephasing of electrons
but leaves their trajectories due to phase space restrains in a 2D
system practically unaffected \cite{ref11}. As long as the
electron-electron scattering mean free path $l_{ee}$ is larger
than the distance between QPCs $L$ we can use a one-collision
approximation to consider this effect ~\cite{ref11}, i.e. an
electron that has been scattered is unlikely to be scattered again
before it reaches the detector. It has been shown \cite{ref11}
that the contribution of electrons that reach the detector
ballistically can be approximated by:

\begin{figure}
\includegraphics[width=8cm]{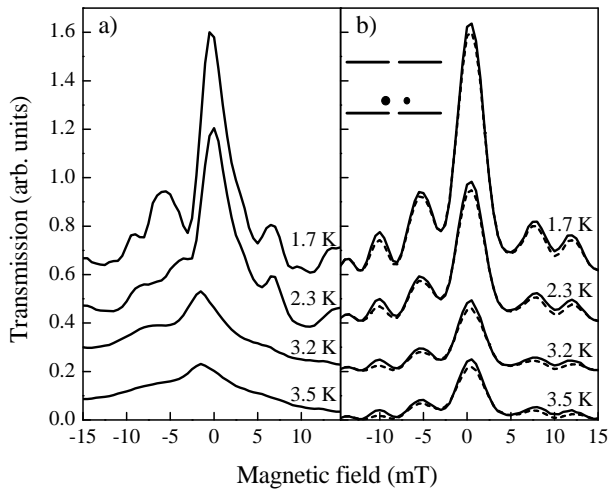}
\caption{\label{fig5} Measured electron beam profiles (a) and
calculated interference patterns (b) at different temperatures
(solid lines: only thermal broadening is considered, dashed lines:
thermal broadening and electron-electron scattering effects are
included). Results are displayed with a constant shift for
clarity. Parameters for the scattering potentials are:
$x_i$=1.2~$\mu$m, $y_i=-0.26~\mu$m, $\Delta r_i=0.08~\mu$m,
$V_{0}\approx-8$~meV for first impurity (dip) and
$x_{i}=1.2~\mu$m, $y_i=0.23~\mu$m, $\Delta r_i=0.04~\mu$m,
$V_{0}\approx82$~meV for second (barrier).}
\end{figure}

\begin{equation}
\Psi_{b}(L,y)=\Psi(L,y)\exp\left(-\frac{L}{l_{ee}}\right) ,
\label{eq-10}
\end{equation}
where $l_{ee}$ is determined using the usual expression for the
energy relaxation length in a 2DEG \cite{ref15}:

\begin{multline}
l_{ee}=v_{F}\tau_{ee},\\
\frac{1}{\tau_{ee}}=-\frac{E_{F}}{4\pi\hbar}{\left(\frac{\Delta}{E_{F}}\right)^{2}}
\left[\ln\left(\frac{\Delta}{E_{F}}\right)-\ln\left(\frac{2~q_{TF}}{k_{F}}\right)-\frac{1}{2}\right],
\label{eq-11}
\end{multline}
where $v_{F}=\sqrt{\frac{2E_{F}}{m^{*}}}$ is the Fermi velocity,
$\tau_{ee}$ is the electron-electron scattering time, $\Delta$ is
the electron excess energy relative to $E_{F}$, $q_{TF}$ is the 2D
Thomas-Fermi screening wave vector, and
$k_{F}=\frac{\sqrt{2m^{*}E_{F}}}{\hbar}$. Electrons that have been
scattered lose their phase information, and reach the detector
with an arbitrary phase. Their contribution to the transmitted
signal can be expressed as:

\begin{equation}
\Psi_{s}(L,y)=\Psi^{0}(L,y)\left[1-\exp\left(-\frac{L}{l_{ee}}\right)\right]
, \label{eq-12}
\end{equation}
where $\Psi^{0}(L,y)$ is the wave transmitted from injector to
detector QPC without impurities in the channel [Eq.~(\ref{eq-1})].
The total propagated signal is then determined as the sum of
Eq.~(\ref{eq-10}) and Eq.~(\ref{eq-12}). The results of this
calculation are presented in Fig.~\ref{fig5}~b) (dashed lines) for
different temperatures from 1.7~K to 3.5~K. It can be seen that
the effect of electron-electron scattering contributes to a
further decrease and smearing of the interference pattern
[Fig.~\ref{fig5}~b), dashed lines]. However, even at 3.5~K the
structure is still more pronounced than experimentally observed.
We conclude that even more dephasing processes contribute to the
experimental data, e.g., electron-phonon scattering processes,
which have not been considered in our model.

\begin{figure}
\includegraphics[width=8cm]{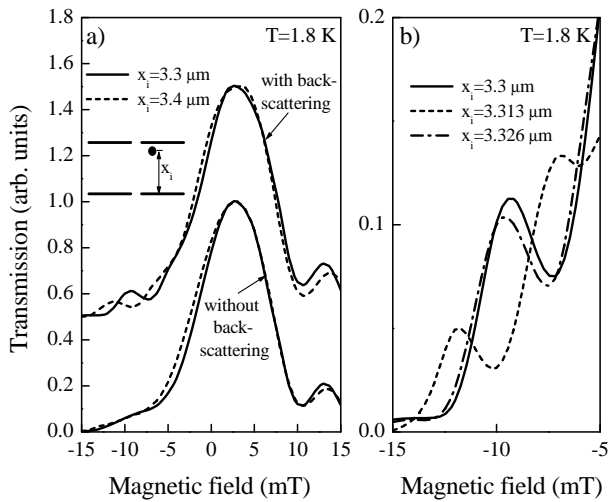}
\caption{\label{fig6} a)~Calculated interference patterns without
and with consideration of the back-scattering effects. Parameters
for the scattering potential are: $x_i=3.3~\mu$m for solid lines
and $x_i=3.4~\mu$m for dashed lines, $y_i=0.15~\mu$m, $\Delta
r_i=0.075~\mu$m, $V_{0}\approx41$~meV (the results including
back-scattering are presented with a shift). b)~Enlarged section
of the upper trace in Fig.~a) for three different values of $x_i$:
$3.3~\mu$m (solid line), $3.313~\mu$m (dashed line), $3.326~\mu$m
(dot-dashed line).}
\end{figure}

Up to now only forward propagating electrons have been considered
in the calculations. However, electrons that are scattered out of
the beam can be scattered again by an additional impurity
potential or the system boundaries and then reach the detector.
For a complete simulation of the device behavior these
back-scattered electrons have also to be considered in the
calculations. Since back-scattered electrons travel much longer
distances before they reach the detector, the effect has to be
taken into account only if the additional path length remains
smaller than the electron-electron scattering length, i.e. smaller
than the phase coherence length. For temperatures considered here,
this holds only for back-scattering effects that occur at
scattering centers located in close vicinity to the injector or
detector, or if two scattering potentials are close to each other.
If the impurity is located near the injector, the back-scattered
wave must be calculated as that part of the injected beam that is
scattered from the impurity towards the injector QPC, reflected at
the injector boundaries, and then retransmitted via the impurity
towards the detector QPC. For this configuration, self-consistent
calculations of the modified injected electron beam $\Psi(0)'$ due
to back-scattered electrons into the injector QPC should be used
in the simulation. We have not attempted to incorporate such
self-consistency into our model. However, a self-consistent method
is not needed when the impurity is located near the detector. In
this case, the back-scattered wave is that part of the beam that
is reflected back from the channel boundaries near the detector
towards the impurity and subsequently is scattered into the
detector QPC. The calculated transmission probabilities without
and with back-scattering effects for a temperature of $T$=1.8~K
are presented in Fig.~\ref{fig6}~a) for the case of an impurity
located near detector QPC. It can be seen that for different
values of $x_{i}$ the interference patterns are very similar when
back-scattering is omitted (lower curves in Fig.~\ref{fig6}~a),
and that the curves differ significantly when back-scattering is
taken into account (upper curves in Fig.~\ref{fig6}~a). To
illustrate this effect more clearly, a part of the upper trace in
Fig.~6~a) is replotted enlarged in Fig.~6 b) (solid line). It can
be seen that if the initial position of the scattering potential
is changed by an amount of the order of $\lambda_F /4$ (Fermi
wavelength $\lambda_F \approx 50$~nm for the samples under
investigation), structures in the electron beam signal which
appeared as a maximum in the initial curve are rendered into a
minimum and vice versa. The initial shape is recovered if the
position is changed by $\approx \lambda_F /2$. This effect, which
obviously results for constructive and destructive interferences,
directly manifests the wave nature of the electron beam.

Our calculations enable us not only to understand electron beam
collimation experiments, but are applicable also to recent
experiments where an electron beam in the vicinity of a QPC has
been manipulated by a charged tip of an atomic force microscope
\cite{ref7,ref16}. The tip of the microscope introduces a
scattering potential which can be located anywhere in the 2DEG
area. Electrons which are back-scattered into the injector are
detected via changes in the injector current. These changes can be
used to map out the current distribution behind the injector QPC.
The effect of coherent constructive and destructive interference
of back-scattered electrons [Fig.~6 b)] is clearly demonstrated in
these experiments. However, in our opinion the authors' conclusion
of a branching of the electron beam around the scattering
potentials is a misinterpretation. An experimental absence of
current variation at the injector QPC does not imply that the
current density is zero at the point where the atomic force
microscope tip is located. Especially when the charged tip is
behind a potential fluctuation of the 2DEG layer no noticeable
changes are expected in the injector current. Based on our
simulation, we surmise that in this situation when the tip and the
scatterer are located close to each other multiple back-scattering
events have to be considered, which drastically reduce the number
of electrons reaching the injector QPC. Therefore, from this
observation no information on the propagating electron beam can be
obtained. Furthermore, a branching of the electron beam can be
excluded due to numerous electron beam collimation experiments
using a second QPC as detector for the actual electron beam
\cite{ref8,ref16,ref17,ref18,ref19} at distances of up to
$4~\mu$m. In none of these experiments 'dark' areas have been
detected. The center part of the beam exhibits always the highest
signal intensity and observed structures are well explained by
wave interference effects. It should be noted that the Frauenhofer
diffraction picture is not sufficient to reproduce the observed
inference patterns in electron beam experiments. A fully quantum
mechanical model is needed, as we have demonstrated here.

\section{\label{sec:level1}Conclusion}

In this paper we presented a comprehensive quantum mechanical
model to calculate the transmission probability of an electron
beam traversing a 2DEG region confined between two opposite
electrostatically defined QPCs. By including local potential
fluctuations, temperature and geometric effects experimental
results can be interpreted in great detail. The main features of
the experimental data are represented in the calculation already
including only one or two locations for a circular shaped
scattering potentials. Even though interpretation of the values
given for the location is sometimes ambiguous due to the symmetry
of the system, the extracted sizes and strengths exhibit only a
few percent of uncertainty and give for the first time reliable
values for the dimensional extents and strengths. It should be
noted here that this symmetry is broken when back-scattering and a
finite mean free path are fully taken into account. The extracted
dimensions and amplitudes of the scattering centers correspond
well with estimations from first principle calculations on the
correlation of remote donors (size: \cite{ref4}, strength:
\cite{ref14}). Also modulation of QPC currents in local probe
experiments can be completely understood in terms of
back-scattered electrons from a scattering potential induced by
the charged local probe.

\begin{acknowledgments}
We acknowledge support by the Alexander von Humboldt foundation,
the German Academic Exchange Service (DAAD), and the Deutsche
Forschungsgemeinschaft (SFB 410).
\end{acknowledgments}



\end{document}